\documentclass[conference,10pt,textwidth=3.33in]{IEEEtran}
\ifCLASSINFOpdf
   \usepackage[pdftex]{graphicx}
   \graphicspath{{../pdf/}{../jpeg/}}
   \DeclareGraphicsExtensions{.pdf,.jpeg,.png}
\else
   \usepackage[dvips]{graphicx}
   \graphicspath{{../eps/}}
\fi
\usepackage[cmex10]{amsmath}
\usepackage{algorithmic}

\usepackage{array}
\usepackage{fixltx2e}

\usepackage{mdwmath}
\usepackage{amsmath}
\usepackage{multirow}
\usepackage{mathtools}

\usepackage{mdwtab}
\usepackage[flushleft]{threeparttable}
\usepackage{stfloats}
\usepackage[ruled,vlined]{algorithm2e}

\usepackage{slashbox}
\usepackage{caption}
\usepackage{subcaption}

\usepackage{threeparttable}

\usepackage{cite}

\makeatletter
 \let\@copyrightspace\relax
\makeatother

\usepackage{epstopdf}

\begin{document}
%
\title{Dynamic Multi-level Privilege Control in Behavior-based Implicit Authentication Systems Leveraging Mobile Devices}

\author{
\IEEEauthorblockN{Yingyuan Yang\IEEEauthorrefmark{1}, Xueli Huang\IEEEauthorrefmark{3}, Yanhui Guo\IEEEauthorrefmark{1}, and Jinyuan Stella Sun\IEEEauthorrefmark{2}\\}
\IEEEauthorblockA{\IEEEauthorrefmark{1}University of Illinois, Springfield, IL, 62629 USA Email: \{yyang260,yguo56\}@uis.edu\\}
\IEEEauthorblockA{\IEEEauthorrefmark{3}Temple University, Philadelphia, PA, 19122 USA Email: tuc36161@temple.edu\\}
\IEEEauthorblockA{\IEEEauthorrefmark{2}University of Tennessee, Knoxville, TN, 37996 USA Email: jysun@utk.edu
}
}


%


\maketitle

\thispagestyle{plain}
\pagestyle{plain}

\begin{abstract}
\boldmath
Implicit authentication (IA) is gaining popularity over recent years due to its use of user behavior as the main input, relieving users from explicit actions such as remembering and entering passwords. However, such convenience comes with a cost of authentication accuracy and delay which we propose to improve in this paper. Authentication accuracy deteriorates as users' behaviors change as a result of mood, age, a change of routine, etc. Current authentication systems handle failed authentication attempts by locking the users out of their mobile devices. It is unsuitable for IA whose accuracy deterioration induces a high false reject rate, rendering the IA system unusable. Furthermore, existing IA systems leverage computationally expensive machine learning, which can introduce a large authentication delay. It is challenging to improve the authentication accuracy of these systems without sacrificing authentication delay. In this paper, we propose a multi-level privilege control (MPC) scheme that dynamically adjusts users' access privilege based on their behavior change. MPC increases the system's confidence in users' legitimacy even when their behaviors deviate from historical data, thus improving authentication accuracy. It is a lightweight feature added to the existing IA schemes that helps avoid frequent and expensive retraining of machine learning models, thus improving authentication delay. We demonstrate that MPC increases authentication accuracy by 18.63\% and reduces authentication delay by 7.02 minutes on average, using a public dataset that contains comprehensive user behavior data.
\end{abstract}


%
\IEEEpeerreviewmaketitle

\section{Introduction}
Rich behavioral data gathered by various sensors embedded in smart devices facilitates the implicit authentication of users based on their behaviors \cite{yang2016personaia,p9,p62}. In general, IA systems authenticate a user by matching her real-time behavior to her historical behavior. Real-time behavior is obtained from one or more sensors whose data can uniquely characterize the user and distinguish her from other users, at the time of authentication. Similarly, historical behavior is obtained from the same sensors in the past and updated after new data is collected. IA schemes typically run at the background and stream data at an appropriate frequency to ensure that data is sufficiently collected and the battery consumption is reasonable. As with any other practical security system, IA systems need to strike a good balance between security and usability. However, it is highly challenging to achieve such balance due to the dynamically changing behaviors of users. On the one hand, we need the system to cope with a user's behavior deviation \cite{yang2016personaia}, e.g., a change of routine, and not falsely reject the user (usability). On the other hand, the system needs to differentiate between a legitimate user's changed behavior and other users' behaviors to prevent falsely accepting adversaries (security). Balancing between such false rejects and false accepts improves the authentication accuracy and is the main focus of this paper.

In addition to the false reject rate, another important measure of system usability is the authentication delay. Authentication delay mainly consists of training delay to obtain historical behaviors and behavior matching delay, which varies across different authentication schemes \cite{khan2014comparative} and is closely relevant to authentication accuracy. The amount and quality of sensor data collected by the system directly affect the authentication accuracy. Insufficient data collection can result in an inferior historical behavior model that is not representative of a user's behavior. Low-quality data can be caused by noisy behavior data (due to either a legitimate user's behavior deviation or adversaries) or noisy sensor readings. Authentication delay is typically increased when the system attempts to improve upon the amount and quality of the collected data since retraining of the machine learning \cite{p67} model and additional data collection are needed. Balancing between authentication accuracy and delay is hence another problem this paper is trying to solve to further enhance usability.

Existing research on IA systems focuses on the effectiveness of IA schemes, i.e., finding suitable behavioral features such as touch, typing, and other motions that uniquely identify users \cite{p62,p7,bo2013silentsense,feng2014tips,shi2011senguard,gurary2016implicit,shahzad2013secure,castelluccia2017towards,p23}. Although authentication accuracy and delay were measured as performance indicators, none of these papers addressed methods to improve them to make the system more user-friendly. We argue that this is a rather important issue to consider since practicality is the key for IA systems to be widely deployed, and provide our solutions in this paper. Specifically, we propose a multi-level privilege control scheme, or MPC, that divides the single privilege level in current IA systems into multiple fine-grained privilege levels. The privilege levels are used to separate apps based on their level of security so that users can still access the less sensitive apps on their smart devices even if their behaviors change. The levels are dynamically adjusted to reflect the user's dynamically changing behaviors, and therefore enhancing the system's authentication accuracy by balancing between false rejects and false accepts. It is challenging to find such a balance because of the difficulty in distinguishing a user's deviated behaviors from other users' behaviors. In other words, a decreased false reject rate may cause an increased false accept rate and vice versa. A fine line needs to be drawn to lower both rates and boost the system's confidence in a user's legitimacy, which requires an in-depth analysis of the existing IA schemes and suitable mathematical modeling. Main contributions of this paper are summarized as follows:

$\bullet$ We propose a multi-level privilege control scheme to address the usability of IA systems, a key issue the existing IA schemes are faced with, by improving the authentication accuracy and delay at the same time. The scheme solves the core problem of how to set and adjust the user's privilege level such that both false reject rate and false accept rate are decreased, where the problem is modeled by applying physical laws that describe the motion of bodies under the influence of a system of forces. To further correct the privilege level adjustment and improve authentication accuracy, we employ a two-factor authentication mechanism in which the secondary factor provides feedback to identify the user's behavior deviation and filter out noisy sensor readings using Kalman filter. The scheme does not rely on additional data collection and adds no authentication delay. The delay is in fact reduced due to the improved authentication accuracy.

$\bullet$ The proposed MPC can be generally applied to most of the current IA systems as long as the output (behavior scores) can be converted to probabilistic values.

$\bullet$ We demonstrate that MPC increases authentication accuracy by 18.63\% and reduces authentication delay by 7.02 minutes on average. The experiments were conducted using a real dataset that contains 130 users' behavior data. The dataset and associated parameter settings are opened to the public, and thus the repeatability is guaranteed.

\section{Preliminary}

\subsection{SVM Classifier \label{psvm}}
SVM is the most widely adopted technique in IA systems \cite{frank2013touchalytics,p65,bo2013silentsense,p62,p66,p67,p35,gascon2014continuous,alzubaidi2016authentication}. Given a training dataset sampled from a group of people, SVM outputs a hyperplane located in a high dimensional space to cluster the data into two classes, the legitimate class and the illegitimate class. During authentication, new data sampled from the current user is verified according to its position in the hyperplane. The user is deemed legitimate if the new data falls in the legitimate class. In the proposed MPC, we need to calculate the distance between the hyperplane and the testing result in a high dimensional space which renders it difficult with SVM's traditional output. Instead, we utilize the probability output calculated by fitting a sigmoid function, $\frac{1}{1+exp(Af_i+B)}$, to the margins of the SVM \cite{platt1999probabilistic}, where $A$ and $B$ are the parameters required to estimate and $f_i$ indicates the margins of the SVM output. The probability output of SVM is called behavior score in this paper. Behavior scores represent users' behaviors in numeric form and are used by the system to deduce users' legitimacy.


\subsection{Kernel Density Estimator}
Kernel density estimator \cite{scott2008kernel,measurekernel,bishop2006pattern} serves as a tool to analyze the usage pattern of the IA system, e.g., legitimate and illegitimate usages in a given time interval, by estimating how often a given behavior score occurs. This is necessary for distinguishing between the legitimate user's deviated behaviors and other (illegitimate) users' behaviors. Kernel density estimator divides the interval into small bins with length $h$, in each of which it calculates the number of behavior scores that fall into the bin. A distribution of the behavior scores is obtained by placing a Gaussian over each score and then adding up the contributions over the whole dataset. The kernel density model is $p(x)=\frac{1}{N}\sum_{n=1}^1\frac{1}{(2\pi h^2)^{D/2}}exp{-\frac{||x-x_n||^2}{2h^2}}$, where $D$ indicates $D-dimensional space$, $N$ is the total number of behavior scores, $x_n$ is the behavior score and $x$ indicates the center of each bin. Kernel density estimator will be used in Section \ref{da} to estimate the occurrence frequency of a particular behavior score.


\subsection{Kalman Filter}
Kalman filter \cite{kalman} is employed in MPC to filter out sensor noise and help correct behavior deviation. The two types of noises it assumes, process noise and observation noise, can be used to model behavior deviation (or behavior noise) and sensor noise, respectively, making it an excellent tool for noise filtering in IA systems. In addition, Kalman filter is loop carried which means it automatically filters out noises at the time of authentication, instead of the need for more data to perform the filtering as in the existing literature \cite{bo2013silentsense,frank2013touchalytics,p65,abramson2013user,draffin2013keysens,lee2015implicit,wang2015understanding}. This property greatly reduces the authentication delay. Kalman filter will be elaborated in Section \ref{da}.



\section{The Proposed MPC Scheme}

We first provide a high-level overview of our MPC scheme before diving into technical details.

\subsection{System Overview}
\begin{figure}[htb]
\vspace*{-0.3cm}
\begin{minipage}{1\linewidth}
\centering
  \begin{subfigure}[a]{\linewidth}
  \hspace*{-0.7cm}
    \centering
    \includegraphics[width=1.099\textwidth,height=0.07\textheight]{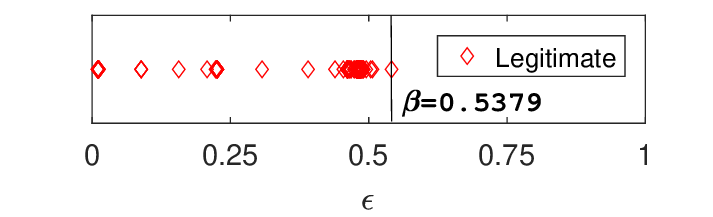}
    \caption{}\label{fig:evaa}
  \end{subfigure}%

  \begin{subfigure}[b]{\linewidth}
  \hspace*{-0.7cm}
    \centering
    \includegraphics[width=1.099\textwidth,height=0.07\textheight]{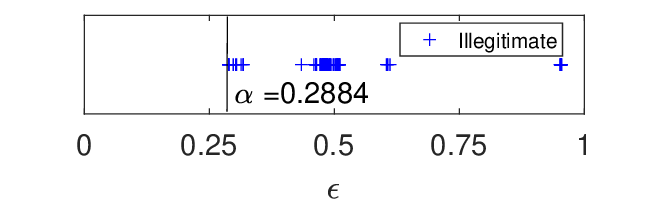}
    \caption{}\label{fig:evab}
  \end{subfigure}%
\end{minipage}%
  \caption{Behavior scores of (a) the legitimate user and (b) the illegitimate users.}
  \label{fig:eva1}
  \vspace*{-0.3cm}
\end{figure}

Existing IA schemes such as \cite{bo2013silentsense,p62} authenticate users by deriving a behavior score using data samples gathered in a period of time, called time window (or authentication cycle) which is a design-specific parameter. This score, $\epsilon$, is then compared with a threshold, e.g., $0.5$. If the threshold is passed, illegitimate usage is indicated and the system will lock the device. When legitimate and illegitimate users have vastly different behaviors, existing IA schemes can achieve high authentication accuracy. However, based on our preliminary experiment using the real dataset \cite{p21,aharony2011social}, more than 70\% of users' behavior data samples overlap and cannot be separated by simply setting a threshold \footnotemark. \footnotetext{The setting and features of the dataset are the same as section \ref{pe}} As an example, we randomly selected two participants from the dataset, one as the legitimate user and the other as the illegitimate user, and converted their SVM output to probabilistic behavior scores. The time window is set to 15 seconds. As shown in Fig. \ref{fig:eva1} (a) and (b), the legitimate and illegitimate users both have a large proportion of behavior scores located around the threshold 0.5 which are inseparable. The behavior overlapping problem can be exacerbated by mimicry attacks where the adversary imitates the legitimate user's behaviors \cite{khan2016targeted}. The MPC scheme attempts to improve authentication accuracy even in the presence of this problem, by using the proposed initial mapping, privilege movement, and domain expansion mechanisms which will be discussed in this section.

\begin{figure}[!h]
\centering
\vspace*{-0.1cm}
\hspace*{-0.2cm}
\includegraphics[width=3.6in,height=1.4in]{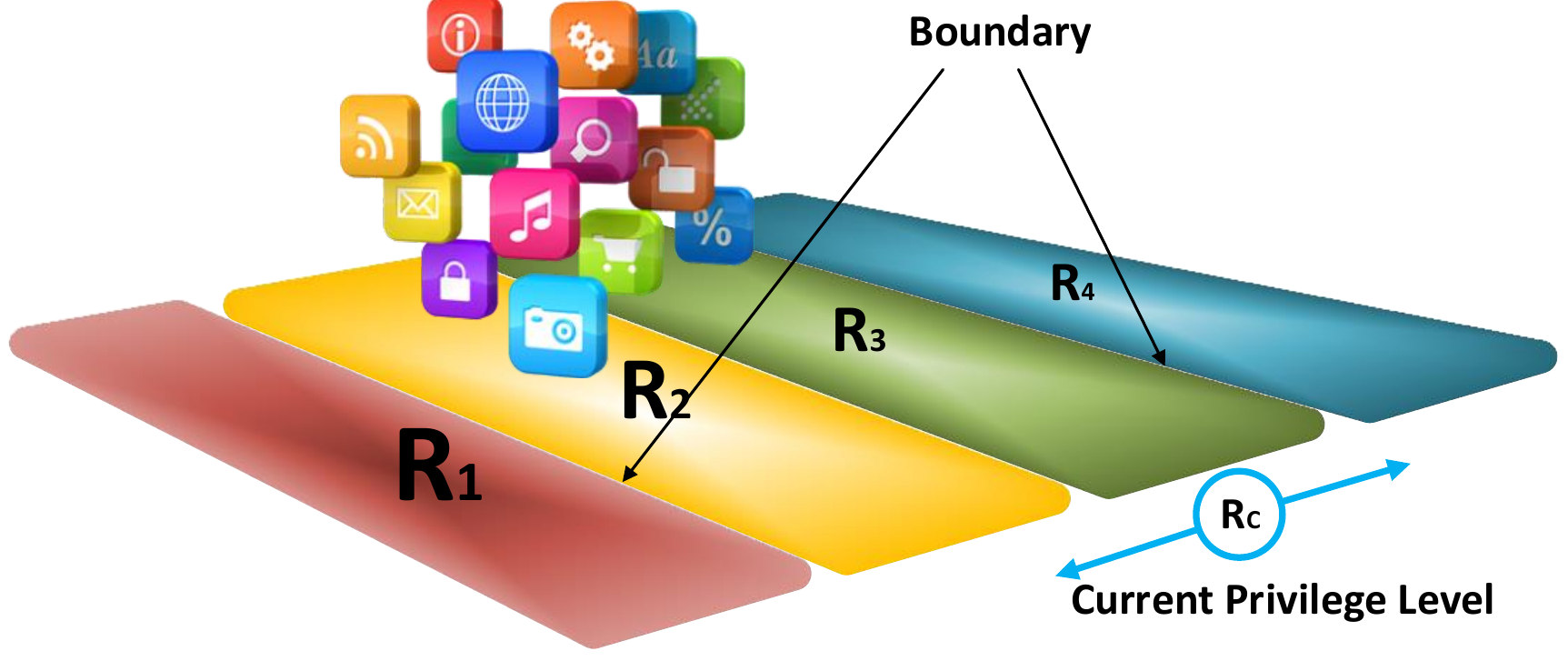}
\caption{The system architecture.}
\label{fig:Sys}
\vspace*{-0.68cm}
\end{figure}

The first step of MPC is to obtain multiple privilege levels using initial mapping. It divides the single privilege level in the existing IA schemes into multiple privilege levels, where each level contains a subset of total installed apps. The system ranks the apps based on their security and risk assessment results provided by \cite{chen2020empirical,ranganath2020free}, and maps them to the privilege levels. Users can later modify the rank after inputting a PIN number, which must be different from the one used to unlock the device. For instance, in a system with four privilege levels $R_1$ through $R_4$, apps can be mapped to the levels as shown in Fig. \ref{fig:Sys}. Apps with the highest security requirements such as banking, e-commerce, health and fitness, credit score, and password manager are mapped to the highest privilege level $R_1$. Apps with lower security requirements such as social media, contacts, games, and utility apps are mapped to lower levels $R_2$ and $R_3$. $R_4$ is the lowest privilege level which corresponds to locking the device and thus contains no app. After obtaining the privilege levels, the system needs to map the user to a specific level $R_c$ based on the user's current behavior at the time of authentication. The level $R_c$ is called the user's current level as shown in Fig. \ref{fig:Sys}. This is performed in the second step of MPC, privilege movement. A user has access to all the apps contained in $R_c$ and the levels below $R_c$, but not the apps in the levels above $R_c$. Moreover, overlapping behaviors are effectively separated in this step. Finally, domain expansion is introduced to dynamically adjust the domain boundaries as more behavior data become available and filter out behavior and sensor noises.

An ideal IA scheme should always map the legitimate user to $R_1$, and illegitimate users to $R_4$. However, as mentioned before, when the legitimate user's behavior deviates, it becomes harder to differentiate it from illegitimate users' behaviors, which is why we need more intermediate levels for such differentiation to reduce the chance of locking the legitimate user. In addition, the intermediate levels also serve as a buffer to prevent the illegitimate user from gaining access to high privacy apps. When behavior deviation happens, instead of locking the device, the legitimate user will be mapped to lower levels and not able to access high-privilege apps. To resume her full access privilege, the user can choose to proactively pass a second-factor authentication, or let the system to adjust her privilege in the next authentication cycle as more data becomes available to make an accurate decision. The two methods are captured in the privilege movement and domain expansion steps. Two-factor authentication has gained increasing popularity and deployment since it enhances security. We use it in our IA scheme with a twist, i.e., instead of having to pass the two factors at the same time, the user will be mapped back to $R_1$ if she passes the second-factor authentication. The reason for such design is that since IA systems are still in their infancy, understanding their performance limitations is the most important first step before we can mature their design. The second factor serves as a feedback mechanism in MPC to help separate behavior deviation from illegitimate behaviors, and fundamentally improve the system's false accept and false reject rates. Despite that we use PIN (different from the password used to unlock the device) input as the second factor in this paper, any authentication scheme other than behavior-based IA can be used. Note that PIN input happens only when there is authentication failure in our system, much less frequently than using a password as the main authentication scheme. After gaining enough insight, we will be able to enhance our IA scheme without the second factor in the future.

The majority of current IA research tends to gather their own data from a small number of volunteers \cite{khan2016targeted,p65,bo2013silentsense,p62}, rendering it difficult to repeat their tests. We use an open dataset \cite{p21} that contains comprehensive user behavior data for the presentation and evaluation of our MPC scheme. Specifically, the dataset contains 130 participants' 8GB data collected in a 5-month period. Data consists of 9 main features: GPS, accelerometer, SMS, app installation, battery usage, call logs, app usage, blue-tooth devices log, and Wi-Fi access points. The details of the features and associated information are shown in \cite{aharony2011social}.

\subsection{Initial Mapping \label{ilm}}
We mainly discuss applying MPC to SVM-based IA schemes. For the other IA schemes \cite{p14,bo2013silentsense,yang2016personaia}, since their output is already a probabilistic behavior score, MPC can be directly applied.

DEFINITION 1. \emph{Let behavior score $\epsilon \in$ [0, 1] denote the probabilistic output of an SVM approximated by a two-parameter sigmoid function $\frac{1}{1+exp(Af_i+B)}$. In a specific training set \footnotemark, we further divide the interval [0, 1] into $n$ sub-intervals, called domains, denoted by $D_n\subset$ [0, 1]. The legitimate domain is the largest sub-interval that contains only true accept (TA) behavior scores. The illegitimate domain is the largest sub-interval that contains only true reject (TR) behavior scores. The slack domain is the sub-interval in between the legitimate domain and illegitimate domain.} \footnotetext{A training set is a dataset that contains various users' historical behavioral data.}

\begin{figure}[htb]
\vspace*{-0.5cm}
\centering
  \begin{subfigure}[b]{.5\linewidth}
  \hspace*{-0.4cm}
    \centering
    \includegraphics[width=2in,height=1.65in]{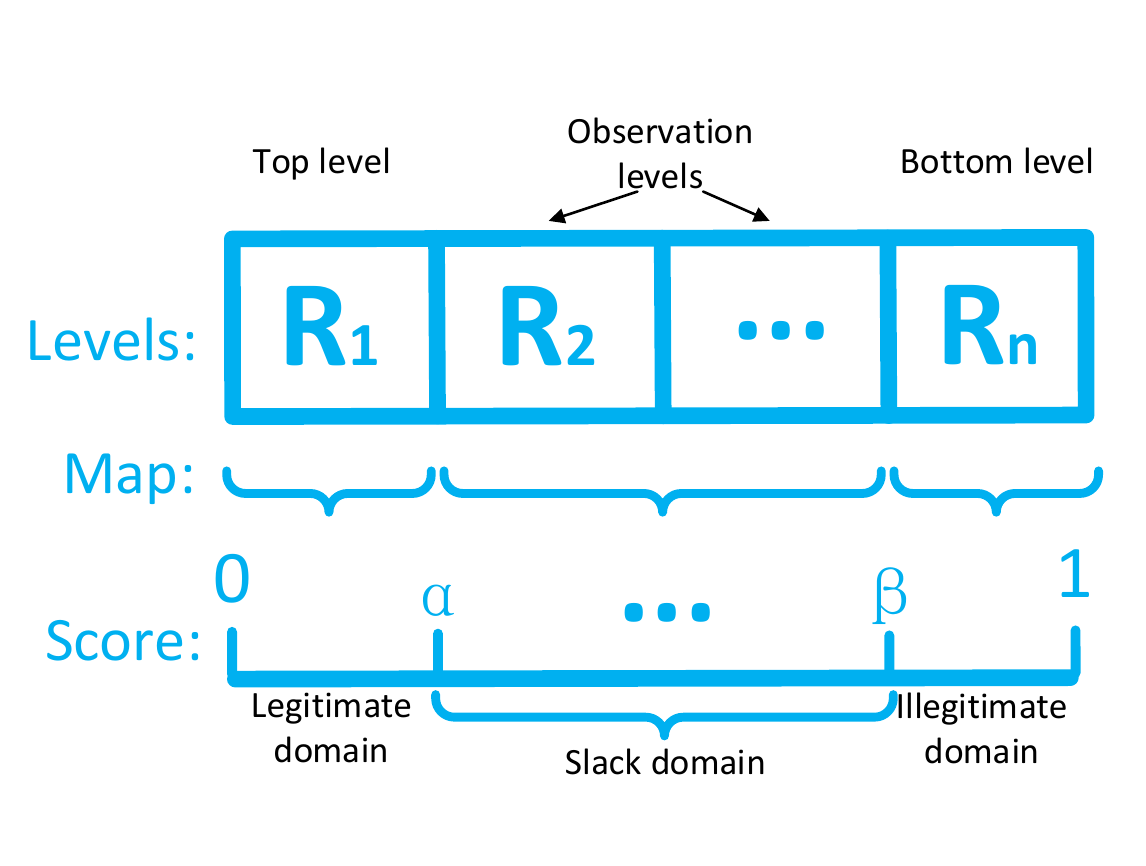}
    \caption{}\label{fig:arca}
  \end{subfigure}%
  \begin{subfigure}[b]{.5\linewidth}
  \hspace*{-0.2cm}
  \vspace*{-0.20cm}
    \centering
    \includegraphics[width=1.9in,height=1.5in]{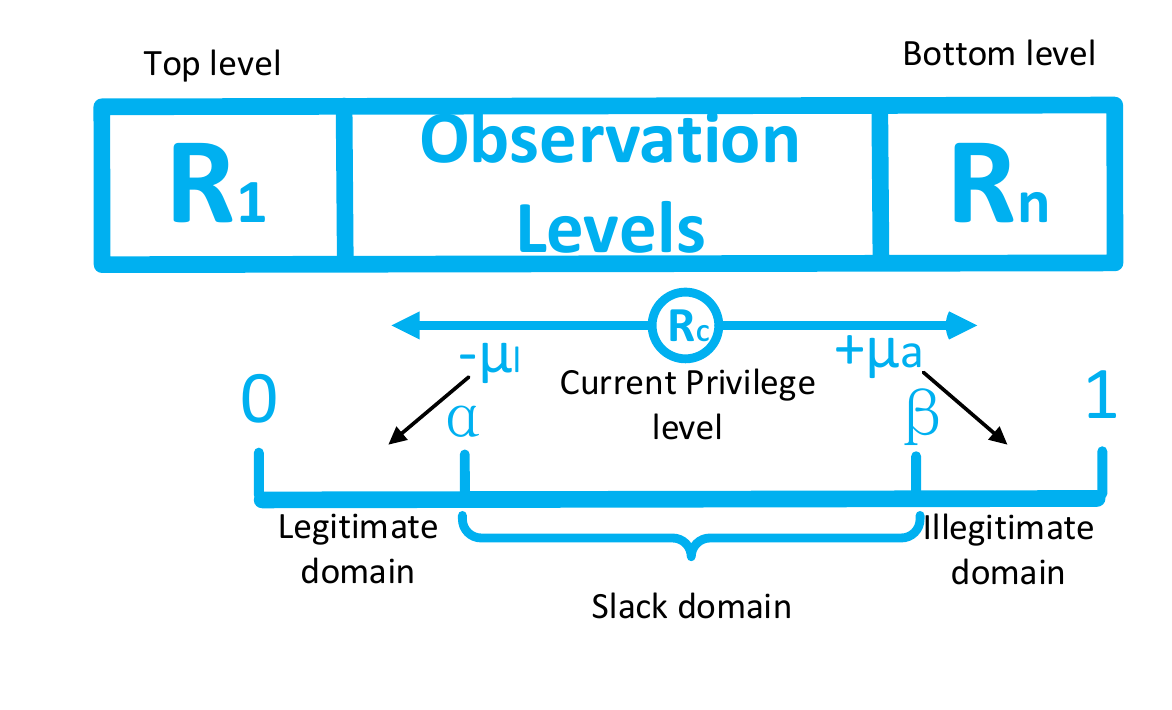}
    \caption{}\label{fig:arcc}
  \end{subfigure}%
  \caption{Initial mapping and privilege movement. (a) Initial mapping. (b) Privilege movement.}
  \label{fig:arc}
\end{figure}

The initial mapping mechanism is illustrated in Fig. \ref{fig:arc} (a). The system first initializes the value of parameters $\alpha$ and $\beta$ by fitting the sigmoid function to the SVM output trained by data sampled from legitimate and illegitimate users. The legitimate and illegitimate domains are predefined based on these two parameters. Assuming the system has $n$ privilege levels, in each authentication cycle as new data is collected, the SVM takes the data as input and outputs a new behavior score indicating the system's authentication decision. If the new score falls in the legitimate domain, the system will move the user's current privilege level $R_c$ to $R_1$ (if $R_c\neq R_1$) which grants the user full access. If the new score falls in the illegitimate domain, the system will lock the device. If the new score falls in the slack domain, the system will map $R_c$ to one of the observation levels $R_2$, $R_3$, ..., $R_{n-1}$, where the user has only limited access.

As shown in Fig. \ref{fig:eva1} (a) and (b), the legitimate and illegitimate domains are [$\beta$, 1] and [0, $\alpha$], respectively. The slack domain is located in [$\alpha$, $\beta$], which contains ambiguous behavior scores that could come from either the legitimate user or illegitimate users and need separation. In a given dataset, we can find $\alpha$ and $\beta$ by searching for the largest and smallest behavior score $\epsilon$ derived from the legitimate user's and illegitimate users' training data, respectively. In this section and the next, we first assume that $\alpha$ and $\beta$ are fixed and focus on the mapping of the current privilege level $R_c$ to one of the observation levels in the slack domain. We then release this assumption in Section \ref{da} when we complete our discussion with the possible movement of the domain boundaries. Compared to the existing IA schemes, the initial mapping in MPC balances between security and usability. Since the system only grants full access to the user who is most likely to be legitimate, security is enhanced. When the likelihood declines, instead of completely locking the user out, the system maps the user to an observation level that grants lower access rights. It enhances usability if the user is legitimate while limiting the security breach if the user is illegitimate. Nevertheless, initial mapping only handles failed authentications in a more gradual way by adding the slack domain and observation levels. It does not fundamentally ameliorate the false reject (FR) and false accept (FA) performance which will be the focus of privilege movement and domain expansion.

\subsection{Privilege Movement \label{dj}}
In initial mapping, the current privilege level $R_c$ is mapped to one of the defined privilege levels $[R_1, R_2, ..., R_n]$ when a new behavior score becomes available at the time of authentication and remains in that level until more data comes in. Such a mapping mechanism does not fundamentally improve the FR and FA performance since the system still needs a way to confirm the user's legitimacy once her behavior score is mapped to the uncertain observation level. Recall that the system's goal is to eventually grant the user full access if she is legitimate and to lock her out otherwise. The slack domain is just a buffer for a smoother transition. We introduce privilege movement in the mapping of $R_c$, where $R_c$ is moved up (towards $R_1$) or down (towards $R_n$) gradually out of the slack domain. We assume it takes the illegitimate users several tries before being able to impersonate (i.e., imitate the behavior or guess the password of) the legitimate user. We also assume that the IA scheme gives high authentication accuracy, i.e., the legitimate and illegitimate users' behavior scores fall into their corresponding domains rather than the slack domain, when the scheme is newly trained. Authentication accuracy will gradually decline as more behavior data becomes available from either the legitimate user or illegitimate users after training. Retraining of the IA scheme may be needed which is covered in detail in \cite{p67}.

\begin{figure*}[htb]
\vspace*{-0.1cm}
%
  \begin{minipage}{.33\textwidth}
  \hspace*{-0.2cm}
  \begin{subfigure}[c]{\linewidth}
    \centering
    \includegraphics[width=1.099\textwidth,height=0.152\textheight]{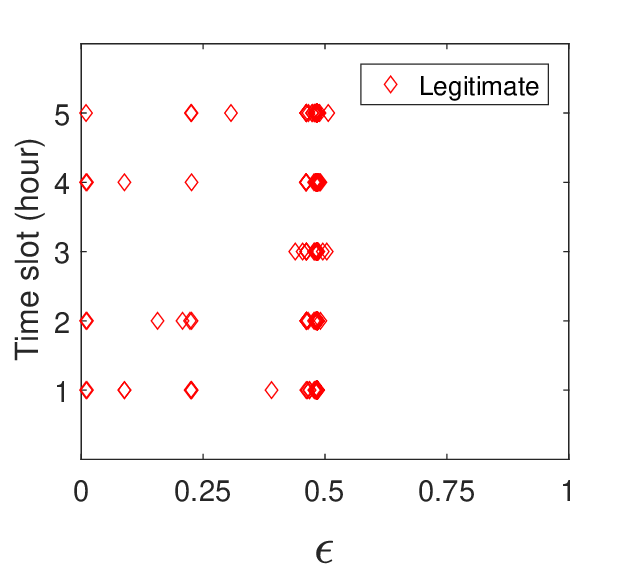}
    \caption{}\label{fig:evac}
  \end{subfigure}%
\end{minipage}%
\begin{minipage}{.33\textwidth}
  \hspace*{-0.2cm}
  \begin{subfigure}[d]{\linewidth}
    \centering
    \includegraphics[width=1.099\textwidth,height=0.152\textheight]{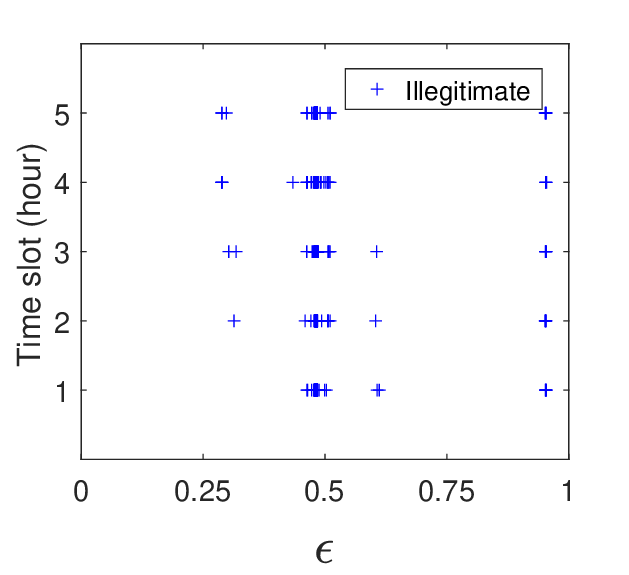}
    \caption{}\label{fig:evad}
  \end{subfigure}%
\end{minipage}%
\begin{minipage}{.33\textwidth}
  \hspace*{-0.2cm}
  \begin{subfigure}[e]{\linewidth}
    \centering
    \includegraphics[width=1.099\textwidth,height=0.152\textheight]{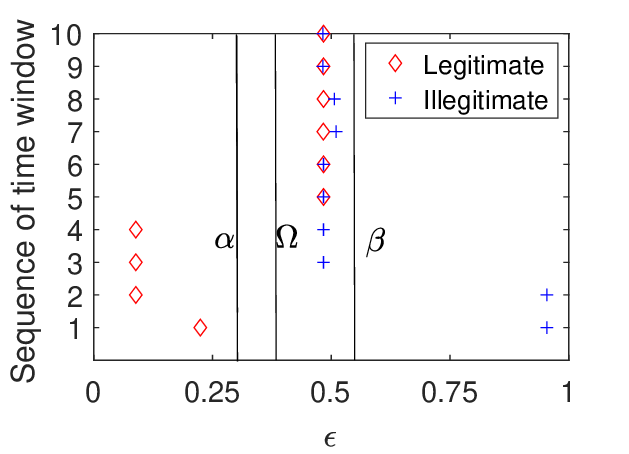}
    \caption{}\label{fig:evae}
  \end{subfigure}%
\end{minipage}%
  \caption{Behavior scores of (a) the legitimate user in a 5-hour period, (b) the illegitimate users in a 5-hour period, and (c) both users in 10 time windows.}
  \label{fig:eva}
  \vspace*{-0.5cm}
\end{figure*}

We summarize the privilege movement mechanism in Fig. \ref{fig:arc} (b). The system keeps track of the user's behaviors and once it observes a behavior score that falls into the slack domain, it searches through the previous scores to find a more definitive answer. If there were scores in the legitimate domain, the system leans towards regarding the user as legitimate and moves $R_c$ upward with distance $-\mu_l$ at the end of the current authentication cycle. This process is repeated until $R_c$ reaches $R_1$. Similarly, if there were scores in the illegitimate domain, the system leans towards regarding the user as illegitimate and moves $R_c$ downward with distance $+\mu_a$ at the end of the current authentication cycle. This process is repeated until $R_c$ reaches $R_n$. If $R_c$ falls in between privilege levels, the user is assumed to access privilege of the lower level. The movement distances $-\mu_l$ and $+\mu_a$ are design parameters that can be constants or variables. For the discussion in this subsection, we let $\mu_l=l/2$ and $\mu_a=l$ where $l$ is the fixed distance between privilege levels. The system is thus less tolerable and more restrictive when there is evidence that the current user is illegitimate. It is also more conservative in giving the user higher access privilege when the user's legitimacy was confirmed in the past but is currently in doubt. This design is to enhance security while not sacrificing usability. Moreover, the FR and FA performance are improved since the system always tries to move $R_c$ out of the slack domain based on evidence. The privilege movement mechanism has $O(1)$ time complexity, which renders the authentication delay the same as the IA schemes without MPC. In the next subsection, we discuss making $\mu_l$ and $\mu_a$ variables to improve authentication accuracy.

As an example, the behavior score distribution for the legitimate user and the illegitimate user is shown in Fig. \ref{fig:eva} (a) and (b), respectively, using the aforementioned experiment with two participants. The scores are grouped into five one-hour time slots, where each time slot contains multiple time windows.
In each time slot, there are behavior scores belonging to the legitimate/illegitimate domain that co-occur with scores belonging to the slack domain. The scores that belong to the legitimate/illegitimate domain are used as evidence and guidance to move the scores in the slack domain. When behavior deviation happens, initial mapping may map the legitimate user to the observation level and still cause false rejects which are corrected with privilege movement. The same is true for false accepts. In addition, we randomly selected a time slot from Fig. \ref{fig:eva} (a) and (b), and magnified it in Fig. \ref{fig:eva} (c) where the threshold $\Omega$ is predefined to best separate the two users. For the ease of presentation, we assume that there is only one observation level and three privilege levels in total. In the first time window, the legitimate user's behavior score falls in the legitimate domain (shown in the figure) but her $R_c$ has not reached $R_1$ (not shown in the figure). The system therefore moves $R_c$ upward for $l/2$. In the second through the fourth time windows, the score falls in the legitimate domain again but $R_c$ has reached $R_1$. So $R_c$ remains in $R_1$. In the fifth through the tenth time windows, $R_c$ falls in the slack domain. Since the system observed four behavior scores in the legitimate domain, $R_c$ remains in $R_1$. If the system observed scores in the illegitimate domain instead, $R_c$ would have been moved towards $R_n$. The illegitimate user in Fig. \ref{fig:eva} (c) follows a similar privilege movement process. Using the dataset, we were able to observe the co-occurrence of legitimate/illegitimate-domain behavior scores and slack-domain behavior scores for the same user in a reasonably short period of time (2-3 minutes), in all of the two-participant experiments we conducted.

The effectiveness of privilege movement is highly dependent on the size of the legitimate and illegitimate domains. If $\alpha$ and $\beta$ are fixed, they may become less indicative as more behavior data from either the legitimate user or illegitimate users become available. This problem will be addressed in the domain expansion mechanism where the size of the domains is dynamically adjusted to reflect the behavior change and improve the authentication accuracy.

\subsection{Domain Expansion \label{da}}
\begin{figure}[!h]
\centering
\includegraphics[width=3.1in,height=1.4in]{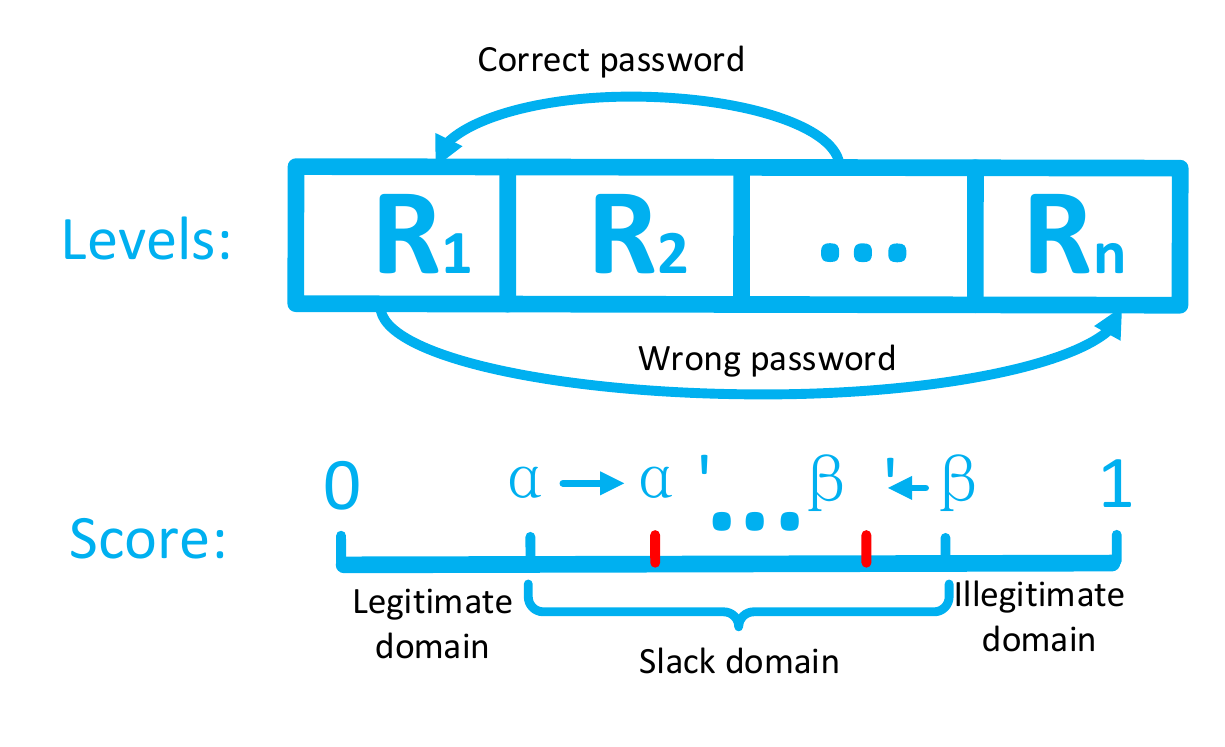}
\caption{Domain expansion.}
\label{fig:dj}
\end{figure}

We now introduce domain expansion in which the domain boundaries $\alpha$ and $\beta$ are updated. In practice, due to behavior deviation and sensor noise, the initial setting of $\alpha$ and $\beta$ may become inaccurate. If behavior scores from the legitimate user keep falling in the slack domain, it may indicate that the legitimate domain is too small and needs to be expanded to reduce false rejects. Similarly, the illegitimate domain may need to be expanded to reduce false accepts. Authentication accuracy is improved as a result. As shown in Fig. \ref{fig:dj}, the original legitimate and illegitimate domains are [0, $\alpha$] and [$\beta$, 1], respectively. The new domains become [0, $\alpha'$] and [$\beta'$, 1] after expansion. In addition, authentication delay is reduced since less privilege movement is needed and the system can make decisions more quickly. In a given dataset, it is straightforward to find out whether the behavior scores that keep falling in the slack domain belong to the legitimate user. In reality, however, it is difficult for the system to know in which case the second-factor authentication (PIN input for our discussion) is needed to provide feedback, as previously mentioned. We assume that the legitimate user will have a large chance to input a correct password at the beginning of usage. Similarly, we assume that the illegitimate user will also have a large chance to input incorrect password at the beginning of usage. Although the illegitimate user can guess passwords, after several unsuccessful tries the chance that the illegitimate user being locked out is increased exponentially due to illegitimate domain expansion.

We model domain expansion by applying physical laws that describe the motion of bodies under the influence of a system of forces. Specifically, the expansion $S$ in time $t$ is defined as:
\begin{equation}
\label{fun3}
\begin{split}
S=\frac{1}{2}(a-\hat{a})t^2+v_0t,
\end{split}
\end{equation}
where $a$ denotes the acceleration of the expansion, $t$ denotes the number of time windows or authentication cycles, $v_0$ denotes the initial velocity of the expansion, and $\hat{a}$ is the resistance that slows down or stops the expansion. Every time the user inputs the correct password and the behavior score is outside of the legitimate domain, the system will expand the legitimate domain to contain the behavior score where the expansion is proportional to the distance between the behavior score and legitimate domain ($\epsilon-\alpha$). However, if the system observes frequent password input, it is an indication that the current machine learning model in the IA scheme is no longer suitable and needs to be retrained \cite{p67}.

The acceleration of the expansion $a$ is defined as:
\begin{equation}
\label{fun1}
\begin{split}
a=\frac{R_d*\varepsilon}{W_1}+W_2+\delta,
\end{split}
\end{equation}
where $W_1=\frac{\sum_i n_l^{(i)}+n_a^{(i)}}{\sum_i N^{(i)}}$ is a balancing parameter that controls the expansion, $W_2$ is a constant representing the initial acceleration, $\sum_i n_l^{(i)}$ is the number of times the user inputs the correct password when her score is in the slack domain, $\sum_i n_a^{(i)}$ is the number of times the user inputs a wrong password when her score is in the slack domain, $\sum_i N^{(i)}$ is the total number of authentication cycles, $R_d$ is the distance between $R_c$ and $R_1$, $\varepsilon=\epsilon-\alpha$ is the distance between the behavior score and legitimate domain, and $\delta$ is the mixture of behavior noise and sensor noise.

The expansion of the legitimate domain may result in the inclusion of illegitimate users' behavior scores that originally fall in the slack domain. To reduce such false accepts, we introduce the resistance $\hat{a}$ that constrains the expansion:
\begin{equation}
\label{fun2}
\begin{split}
\hat{a}=a(\int_0^\alpha p(\varepsilon_a)d\varepsilon_a +\theta),
\end{split}
\end{equation}
where $\theta$ is a constant that prevents $\alpha$ from surpassing $\beta$, $\int_0^\alpha p(\varepsilon_a)d\varepsilon_a$ denotes the probability that the legitimate domain contains behavior scores derived from illegitimate users in the training set, and $\varepsilon_a$ denotes the behavior score derived from illegitimate users' data in the training set. $\int_0^\alpha p(\varepsilon_a)d\varepsilon_a$ is estimated using kernel density estimator.

Substituting (\ref{fun2}) into (\ref{fun3}) and assuming $t=1$, we have
\begin{equation}
\label{fun4}
\begin{split}
S=\frac{1}{2}a(1-\int_0^\alpha p(\varepsilon_a)d\varepsilon_a-\theta)+v_0,
\end{split}
\end{equation}
where $V=1-\int_0^\alpha p(\varepsilon_a)d\varepsilon_a-\theta$ controls when the expansion stops.

Substituting \ref{fun1} into \ref{fun4}, we have
\begin{equation}
\label{fun5}
\begin{split}
S=\frac{1}{2}(\frac{R_d*\varepsilon}{W_1}+W_2)V+v_0+\Delta,
\end{split}
\end{equation}
where $\Delta=\frac{V*\delta}{2}$ is estimated and eliminated using a Kalman filter.

In each authentication cycle, if the user inputs the correct password, the predicted state estimate $x_{k|k-1}$ which controls the expansion of the legitimate domain is defined as: $x_{k|k-1}=F_kx_{k-1|k-1}+B_ku_k$, where
$F_k=\begin{bmatrix}
    1 &t \\
    0 &1 \\
\end{bmatrix}
$, $B_k=\begin{bmatrix}
    \frac{t^2}{2}\\
    t \\
\end{bmatrix}$ and $u_k=(\frac{R_d*\varepsilon_a}{W_1}+W_2)V$. The predicted estimate covariance $P_{k|k-1}$ is defined as: $P_{k|k-1}=F_kP_{k-1|k-1}F_k^T+Q_k$, where the process noise covariance is $Q_k=\begin{bmatrix}
    \frac{t^4}{4} &\frac{t^3}{2} \\
    \frac{t^3}{2} &t^2 \\
\end{bmatrix}*\sigma_a^2$ with $\sigma_a$ being the magnitude of the process noise (behavior noise).  The innovation covariance is $S_k=H_kP_{k|k-1}H_k^T+R_k$, where $H_k=\begin{bmatrix}
    1 \\
    0 \\
\end{bmatrix}$ and $R_k$ is the covariance of the observation noise (sensor noise). Kalman gain is calculated as: $K_k=P_{k|k-1}H_k^TS_k^{-1}$. Since a Kalman filter is loop carried, we update the state estimate and associated covariance at the end of each authentication cycle as: $x_{k|k}=x_{k|k-1}+K_k(z_k-H_kx_{k|k-1})$, and $P_{k|k}=(I-K_kH_k)P_{k|k-1}$. We calculate the expansion as $P_{k|k}H_k$ and need to rescale it before applying it to real systems.

If the user inputs a wrong password, we let $u_k=\frac{\varepsilon_l}{R_d*W_1}+W_2$, and a similar process happens for the expansion of the illegitimate domain.

In addition to causing false accepts, the expansion of the legitimate domain also affects privilege movement, or more specifically, the distance of the movement $-\mu_l$ and $+\mu_a$. Now that the domain boundaries $\alpha$ and $\beta$ are dynamically adjustable, the distance of the movement needs to be adjusted accordingly. We let $-\mu_l=-\mu_l\frac{\int_0^\alpha p(\varepsilon_l)d\varepsilon_l}{\int_0^\alpha p(\varepsilon_a)d\varepsilon_a}$ and $+\mu_a=+\mu_a\frac{\int_\beta^1 p(\varepsilon_a)d\varepsilon_a}{\int_\beta^1 p(\varepsilon_l)d\varepsilon_l}$, where $\varepsilon_l$ and $\varepsilon_a$ denote the behavior scores derived from the legitimate user's and illegitimate users' data in the training set, respectively, $\int_0^\alpha p(\varepsilon_l)d\varepsilon_l$ and $\int_0^\alpha p(\varepsilon_a)d\varepsilon_a$ denote the probabilities that the legitimate domain contains behavior scores derived from the legitimate user's and illegitimate users' data in the training set, respectively, and $\int_\beta^1 p(\varepsilon_l)d\varepsilon_l$ and $\int_\beta^1 p(\varepsilon_a)d\varepsilon_a$ denote the probabilities that the illegitimate domain contains behavior scores derived from the legitimate user's and illegitimate users' data in the training set, respectively. If the ratio $\frac{\int_0^\alpha p(\varepsilon_l)d\varepsilon_l}{\int_0^\alpha p(\varepsilon_a)d\varepsilon_a}$ is large, it indicates that the legitimate user's behavior scores still dominate the legitimate domain, and the distance of the privilege movement is appropriate. Otherwise, the distance needs to be adjusted.


\section{Performance Evaluation \label{pe}}
We have conducted a comprehensive performance evaluation on a four-level MPC scheme that contains a top level, two observation levels, and a bottom level. Most of the experiments use data from all 130 participants in five months, where we randomly select one participant as the legitimate user and mix her data with the data sampled from all other participants. The experiments are performed 130 times for each participant against all the other participants and averaged results are derived for each test. We keep the illegitimate users' data portion in the range of 50\% to 80\% to simulate a more hostile environment. Features used in the evaluation include GPS, app installation, Bluetooth, and battery usage. GPS contains longitude and latitude. App installation contains app package, installed apps, uninstalled apps, running apps, and their corresponding class names. Bluetooth usage contains the participants' IDs, address, and duration. Battery usage contains health percentage, battery level, voltage, plug information, and brand information. The details of the features and associated information are shown in \cite{aharony2011social}. The features are automatically selected using Wind Vane algorithm \cite{yang2017energy} to best represent the behavioral pattern of different users.

\subsection{Authentication Accuracy}
\begin{figure}[htb]
\centering
  \begin{subfigure}[b]{.5\linewidth}
  \hspace*{-0.2cm}
    \centering
    \includegraphics[width=1.099\textwidth,height=0.13\textheight]{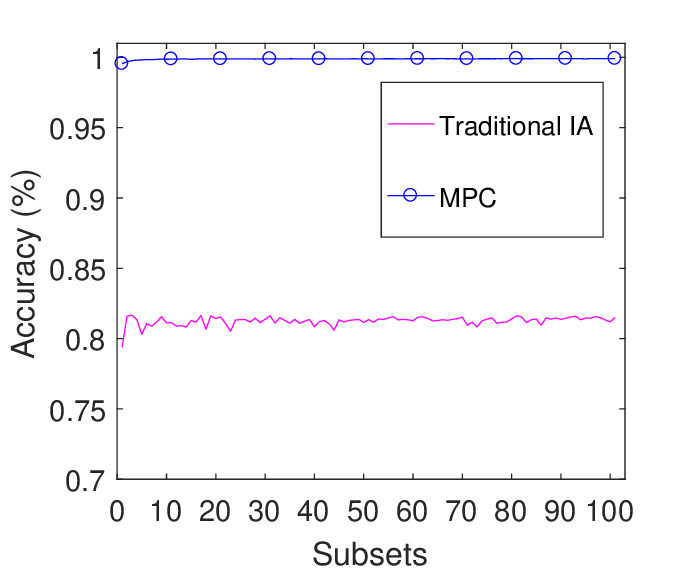}
    \caption{}\label{fig:acca}
  \end{subfigure}%
  \begin{subfigure}[b]{.5\linewidth}
  \hspace*{-0.2cm}
    \centering
    \includegraphics[width=1.1\textwidth,height=0.13\textheight]{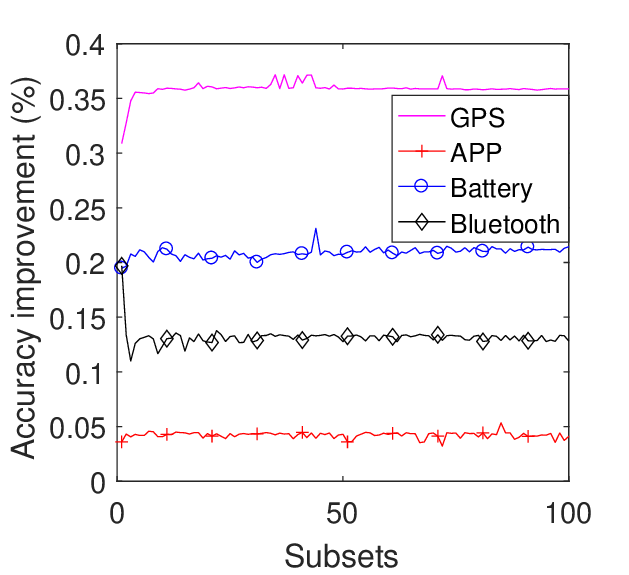}
    \caption{}\label{fig:accb}
  \end{subfigure}%
  \caption{Authentication accuracy. (a) Average accuracy for both MPC and the traditional IA. (b) Accuracy improvement for all four features with MPC.}
  \label{fig:acc}
  \vspace{-0.4cm}
\end{figure}

The time window is set to 15 seconds which contains 1 KB user data. The data is sent to SVM for training and the output is converted to probabilities. To simulate real usage, we divide the whole dataset into 100 distinct subsets sorted based on time, and perform tests by gradually sending the subsets to the system. We use k-fold cross-validation in training and testing. The parameters for SVM, including the separation threshold, are chosen to minimize false rejects and false accepts. Finally, we evaluate the average authentication accuracy in each time window for both MPC (applied to IA) and the traditional IA among all users. We use SVM with the separation threshold $\Omega$ for the traditional IA. The results are shown in Fig. \ref{fig:acc}. In Fig. \ref{fig:acc} (a), the accuracies for both MPC and the traditional IA have some fluctuations in the first five subsets and become stable in the remaining subsets. The accuracy fluctuations of the traditional IA is larger than MPC due to behavior deviation and sensor noise that are left untreated. The accuracy improvement in MPC is obvious and stable across all the subsets. In addition, we calculate MPC's accuracy across all the subsets, which on average achieves 18.63\% improvement compared with the traditional IA.

Fig. \ref{fig:acc} (b) provides a more detailed view of accuracy improvement in the four features. The improvement in GPS is the highest due to the traditional IA's low accuracy using the GPS feature. Similarly, the improvement in app installation is the lowest. The accuracy improvement for MPC is not stable during the first few authentications because of domain expansion. It becomes stable after the 70th subset for all four features. Generally, the accuracy improvement after applying MPC is between 0.04\% to 0.35\%.

\subsection{Authentication Delay}
\begin{figure}[htb]
\vspace{-0.4cm}
\centering
  \begin{subfigure}[b]{.5\linewidth}
  \hspace*{-0.2cm}
    \centering
    \includegraphics[width=1.099\textwidth,height=0.11\textheight]{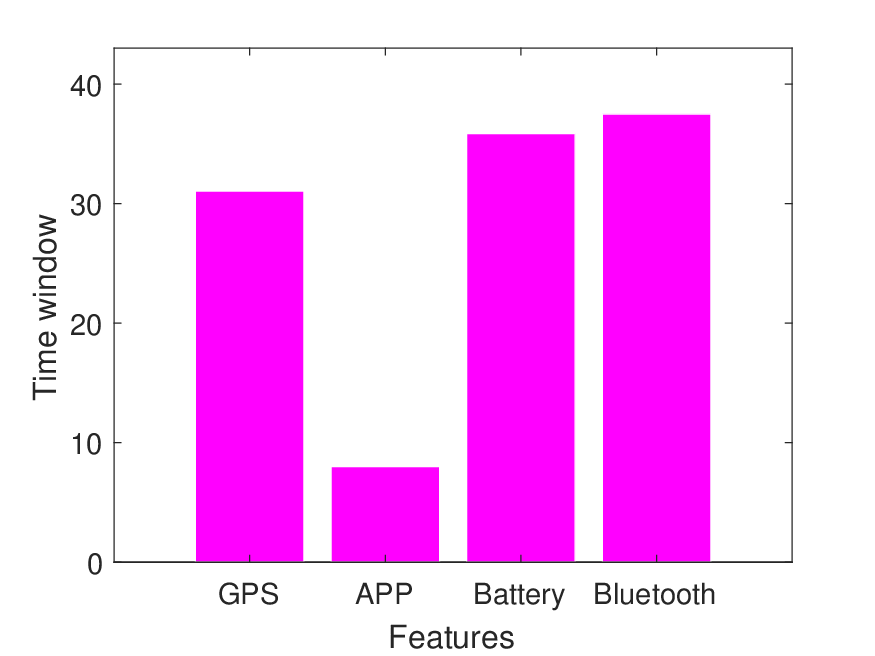}
    \caption{}\label{fig:tra}
  \end{subfigure}%
  \begin{subfigure}[b]{.5\linewidth}
    \centering
    \includegraphics[width=1.1\textwidth,height=0.11\textheight]{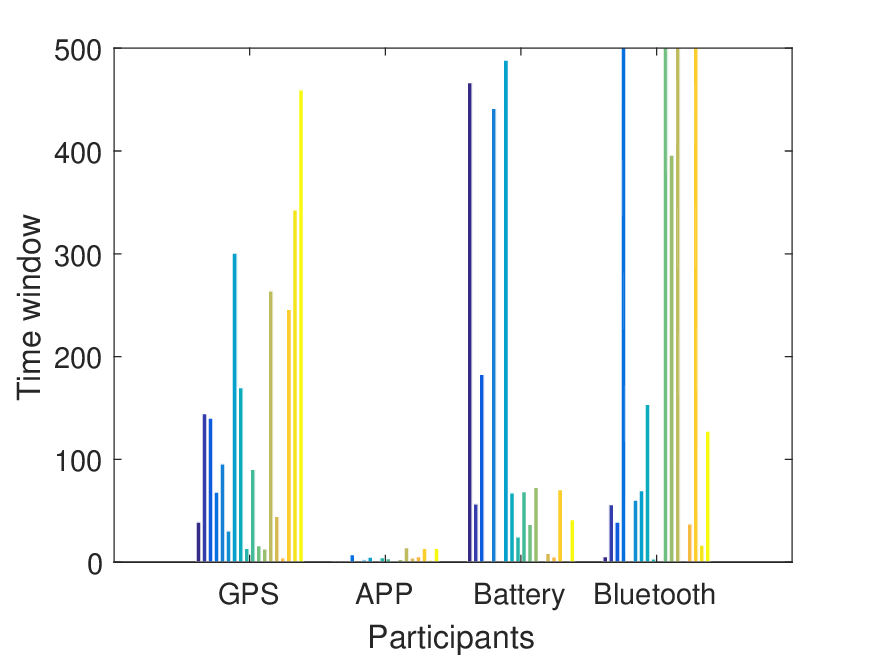}
    \caption{}\label{fig:trb}
  \end{subfigure}%
  \caption{Authentication delay. (a) Average delay reduction for all users. (b) Delay reduction for 18 randomly selected users.}
  \label{fig:tr}
\end{figure}

We calculate the authentication delay for both MPC and the traditional IA when they reach their corresponding highest accuracy, i.e., 99\% and 82\% on average, shown in Fig. \ref{fig:tr}. The average authentication delay reductions for the four features with MPC are shown in Fig. \ref{fig:tr} (a). For the app installation feature, MPC gives the lowest delay reduction since the traditional IA trained by the app data already achieves a good delay performance and there is little room for improvement. Similarly, for the Bluetooth feature, MPC provides the highest delay reduction. In addition, we calculate the overall delay among all features with MPC, which on average achieves a reduction of 7.02 minutes.

Moreover, we calculate the amount of time reduction for each user in MPC when it reaches the highest accuracy, by randomly selecting 18 users, as shown in Fig. \ref{fig:tr} (b). We cluster the results based on the features. Users are marked by different colors and the corresponding delay reduction is shown in the y-axis. It is observed from the results that the delay reduction varies greatly among users and even for the same user, which reflects the complexity of human behaviors.

\subsection{Performance under Long-term Usage}
We evaluate the performance of MPC under long-term usage using data from 19 randomly selected users in three time slots containing 200, 300, and 500 time windows. We calculate the accuracy (ACC), precision (PREC), true accept rate (TAR), true reject rate (TRR), false accept rate (FAR) and false reject rate (FRR) in Table \ref{tbperformance} for both MPC and the traditional IA. As shown in Table \ref{tbperformance}, the performance improvement with MPC is significant compared with the traditional IA. Another important observation is that the performance of the traditional IA does not monotonically increase with time. In other words, the authentication accuracy of the traditional IA does not always improve as we gather more behavior data. This is due to behavior deviation and sensor noise. MPC, on the other hand, is much more predictable in terms of improving the authentication accuracy since it automatically corrects behavior deviation and filters out noise in each authentication cycle.

Furthermore, as shown in Table \ref{tbperformance}, MPC's accuracy improvement becomes smaller between the 300 and 500 time windows, compared with between the 200 and 300 time windows. As discussed previously, MPC reduces the impact of overlapping behavior scores in the slack domain using initial mapping, privilege movement, and domain expansion. Since it is loop carried, the accuracy improvement is reflected gradually in each time window and the expansion becomes slower and more stable with time.

\begin{table}[!ht]
\footnotesize
\renewcommand{\arraystretch}{1.3}
\caption{Performance evaluation under long-term usage.}
\centering
\begin{center}
\hspace{-0.28cm}
\begin{tabular}{|c|c|c|c|c|c|c|}
\hline
\multicolumn{7}{ |c| }{Traditional IA ($\pm 1.0$)*} \\
\hline
Time&ACC \%&PREC \%&TAR \%&TRR \%&FAR \%&FRR \%\\
\hline
200 &87.58&92.04&89.18&69.38&30.62&10.82\\
\hline
300 &84.40&87.77&87.08&67.84&32.16&12.92\\
\hline
500 &83.16&86.90&84.62&66.77&33.23&15.38\\
\hline
\hline
\multicolumn{7}{ |c| }{MPC ($\pm 1.0$)} \\
\hline
200 &97.26&97.40&98.80&93.52&6.48&1.20\\
\hline
300 &98.64&98.87&98.93&98.18&1.82&1.07\\
\hline
500 &98.97&99.08&99.15&98.72&1.28&0.85\\
\hline
\end{tabular}
\end{center}
\begin{tablenotes}
      \footnotesize
      \item *Time stands for time window. $ACC=\frac{TA+TR}{TA+TR+FA+FR}$, $PREC=\frac{TA}{TA+FA}$, $TAR=\frac{TA}{TA+FR}$, $TRR=\frac{TR}{TR+FA}$, $FAR=\frac{FA}{FA+TR}$ and $FRR=\frac{FR}{FR+TA}$.
\end{tablenotes}
\label{tbperformance}
\vspace{-0.57cm}
\end{table}

\subsection{Other Performance Measures}
\begin{figure}[htb]
\vspace*{-0.7cm}
\centering
  \begin{subfigure}[b]{.5\linewidth}
  \hspace*{-0.2cm}
    \centering
    \includegraphics[width=1.099\textwidth,height=0.13\textheight]{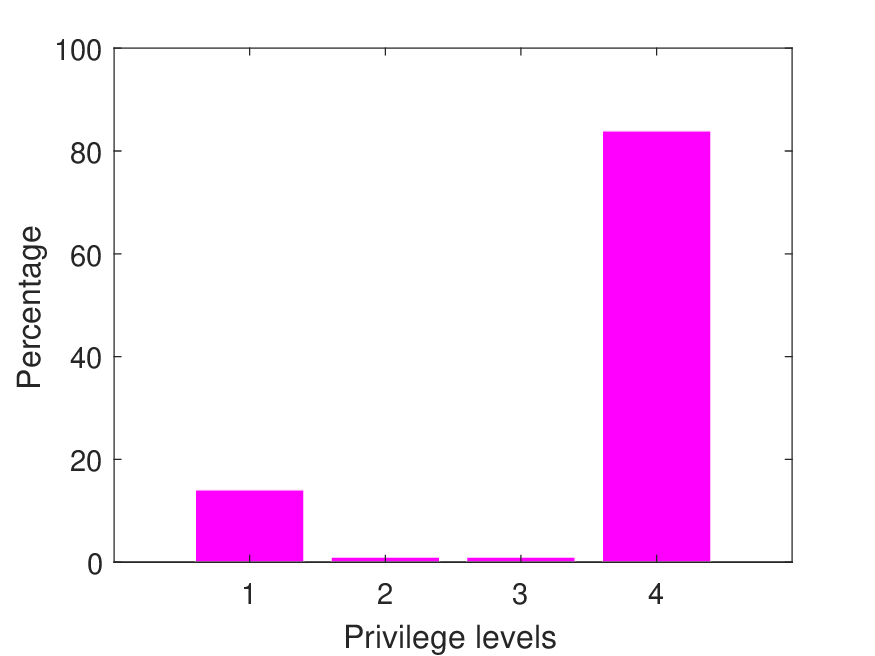}
    \caption{}\label{fig:tra}
  \end{subfigure}%
  \begin{subfigure}[b]{.5\linewidth}
    \centering
    \includegraphics[width=1.1\textwidth,height=0.14\textheight]{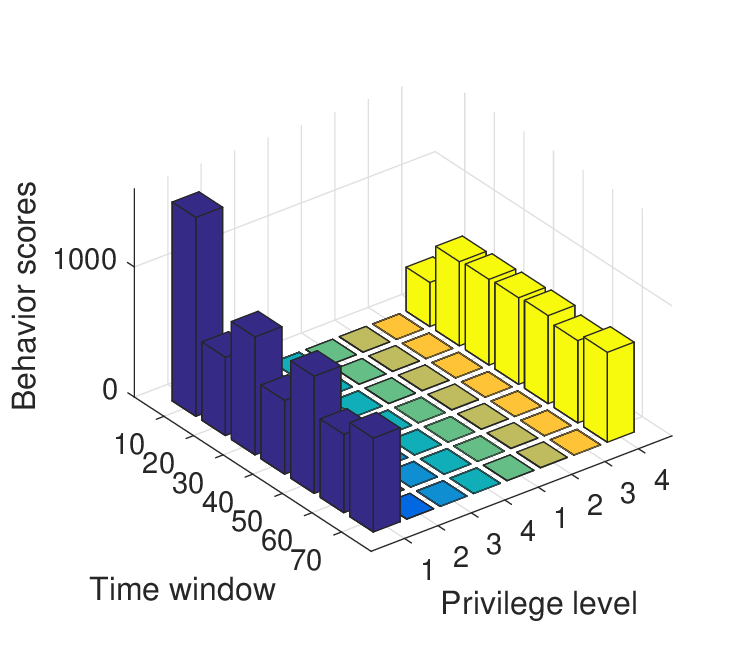}
    \caption{}\label{fig:trb}
  \end{subfigure}%
  \caption{The proportion of behavior scores in each level. (a) Average proportion. (b) The proportion in time windows 10 through 70.}
  \label{fig:levelbin}
  \vspace{-0.47cm}
\end{figure}

We calculate the percentage of behavior scores that are mapped to each privilege level in MPC, for the legitimate user and illegitimate users. As shown in Fig. \ref{fig:levelbin} (a), less than 3\% of the behavior scores are mapped to the observation levels, which indicates that MPC is fast and highly effective in making the final decision. The number of behavior scores that fall in the observation levels, level 2, and level 3, is almost identical.

We also calculate the behavior score distributions in each privilege level for time windows 10 through 70 as shown in Fig. \ref{fig:levelbin} (b). The z-axis indicates the number of behavior scores. The y-axis indicates the time windows. The x-axis indicates the privilege levels, where the left four levels are plotted from the legitimate user's behavior scores, and the right four levels are plotted from illegitimate users. For both users, the number of scores that fall in the observation levels is small, less than 10, which is similar to the result in Fig. \ref{fig:levelbin} (a). The proportion of the behavior scores in the top and bottom privilege levels is different from Fig. \ref{fig:levelbin} (a), since Fig. \ref{fig:levelbin} (b) only considers limited time windows while Fig. \ref{fig:levelbin} (a) considers data spanning 5 months.

\section{Related Work}
The majority of the existing work \cite{p62,p7,bo2013silentsense,feng2014tips,shi2011senguard,shahzad2013secure,castelluccia2017towards} focuses on finding suitable behavioral features such as touch, typing, and other motions that uniquely identify users, ignoring the usability of the IA systems. The amount of data gathered by various sensors directly affects the accuracy of IA systems \cite{p62,yang2016personaia,p9}. By increasing the time spent in collecting users' behavior data, the accuracy of IA can be improved \cite{p62,p9}. However, this approach will also increase the authentication delay and undermine usability. In addition, Hayashi et al. proved that the traditional IA with two privilege levels (locking or unlocking) constrains its performance \cite{hayashi2012goldilocks}. In practice, a more sophisticated system is needed. In this paper, we proposed a multi-level privilege control scheme to address the usability of IA systems by improving authentication accuracy and delay.

To deal with the behavior and sensor noises, most of the existing IA schemes use simple approaches such as resampling \cite{p62}, averaging the results \cite{bo2013silentsense}, or no approach at all \cite{p65,p35,kate2017authentication}. Such noises will degrade system performance in terms of authentication accuracy and delay. The problem will be exacerbated as the size of the behavior data grows. We applied a Kalman filter \cite{welch1995introduction} to correct behavior deviation and filter out sensor noise during the authentication. We showed that a Kalman filter is naturally suitable for IA and can be implemented in practice to further improve authentication accuracy while reducing authentication delay.

The existing IA systems are evaluated using private datasets collected from their volunteers \cite{p62,khan2016targeted,frank2013touchalytics,bo2013silentsense}. Such datasets may not be sharable due to the sensitivity of human behavior data. It is hence difficult to recreate their experiments, compare with their schemes, or use their datasets for future research. Our proposed MPC is evaluated using a public and comprehensive dataset \cite{p21,aharony2011social}. The repeatability of our experiments and fair comparisons are guaranteed.


\section{Conclusion and Future Work}
In this paper, we proposed a multi-level privilege control scheme, MPC, to enhance the usability of IA system. As a framework seamlessly above it, MPC can significantly boost the performance of the original IA system. In MPC, we modeled the privilege changing process of users and bridged the privilege control mechanism to implicit authentication. To this end, we introduced Initial Mapping, Privilege Movement, and Domain Expansion techniques. We evaluated MPC using a public dataset, which on average achieves 18.63\% accuracy improvement and 7.02-minute authentication delay reduction. MPC is a lightweight solution that can be generally applied to IA schemes whose output can be converted to probabilistic values. In the future, to benefit associated research, we will share the system's source code, parameter setting, and dataset on our lab website.

\section{Acknowledgement}
This work was partially supported by the US National Science Foundation (NSF) under grant CNS-1422665 and the Army Research Office (ARO) under grant 66270-CS.

\bibliographystyle{IEEEtran}
\bibliography{reference}
%

\end{document}